\documentclass[%
 reprint,
 superscriptaddress,
 amsmath,amssymb,
 aps,
 pra,
]{revtex4-1}

\usepackage{graphicx}% Include figure files
\usepackage{dcolumn}% Align table columns on decimal point
\usepackage{bm}% bold math
\usepackage{hyperref}% add hypertext capabilities
\begin{document}

\preprint{APS/123-QED}

\title{Critical Entanglement Dynamics at Dynamical Quantum Phase Transitions}

\author{Kaiyuan Cao}
 \email{kycao@yzu.edu.cn}
\affiliation{College of Physics Science and Technology, Yangzhou University, Yangzhou 225009, China}

\author{Mingzhi Li}
\affiliation{College of Physics Science and Technology, Yangzhou University, Yangzhou 225009, China}

\author{Xiang-Ping Jiang}
\affiliation{School of Physics, Hangzhou Normal University, Hangzhou, Zhejiang 311121, China}

\author{Shu Chen}
 \email{schen@iphy.ac.cn}
\affiliation{Beijing National Laboratory for Condensed Matter Physics, Institute of Physics, Chinese Academy of Sciences, Beijing 100190, China}

\author{Jian Wang}
 \email{phcwj@hotmail.com}
\affiliation{College of Physics Science and Technology, Yangzhou University, Yangzhou 225009, China}

\date{\today}

\begin{abstract}
  We investigate the critical behavior of momentum-space entanglement entropy at dynamical quantum phase transitions (DQPTs) in translationally invariant two-band insulators and superconductors. By analyzing the Su–Schrieffer–Heeger model, the quantum XY chain, and the Haldane model, we establish that the geometric DQPT condition $\hat{\textbf{d}}_{\textbf{k}}^{i} \cdot \hat{\textbf{d}}_{\textbf{k}}^{f} = 0$ manifests as exact degeneracy $p_{\textbf{k}^{*}}=1/2$ in the entanglement spectrum defined with respect to the post-quench eigenbasis, yielding a maximal momentum-space entropy of $\ln 2$. In one dimension, critical momenta appear as isolated points, whereas in two dimensions they form continuous one-dimensional manifolds, reflecting the dimensional dependence of the underlying critical structure. Importantly, alternative bipartitions such as the sublattice basis produce qualitatively different behavior: the entropy becomes explicitly time-dependent and attains a minimum at DQPT critical times, underscoring the essential role of basis selection. Our results establish that momentum-space entanglement entropy, when evaluated in the appropriate eigenbasis, provides a robust, time-independent diagnostic of DQPTs and offers a unified geometric perspective linking entanglement, topology, and non-equilibrium criticality.
\end{abstract}

\maketitle

%\tableofcontents

\section{Introduction}
\label{sec: intro}

Quantum information theory has fundamentally reshaped the understanding of quantum phase transitions, with entanglement measures---most notably entanglement entropy---serving as powerful probes of quantum ground states \cite{Amico2008rmp, Eisert2010rmp}. In equilibrium, entanglement entropy not only detects criticality through logarithmic violations of the area law, but also reveals topological order via subleading contributions \cite{Kitaev2006prl, Levin2006prl} and identifies the central charge of the underlying conformal field theory \cite{Calabrese2004jsmte, Vidal2003prl}. Beyond static correlations, out-of-time-ordered correlators (OTOCs) have emerged as complementary diagnostics of quantum criticality, capturing dynamical information scrambling near phase transitions \cite{Wei2019prb} and exhibiting universal scaling governed by critical exponents \cite{Tripathy2026}. Collectively, these information-theoretic quantities have elevated entanglement and scrambling to the status of fundamental order parameters, characterizing phases and transitions beyond the conventional Landau paradigm.

The notion of a phase transition, traditionally tied to singularities in the equilibrium free energy, has been extended to the time domain through the framework of dynamical quantum phase transitions (DQPTs). Central to this framework is the Loschmidt echo---the overlap between an initial state and its time-evolved counterpart---whose rate function develops nonanalytic cusps at critical times \cite{Heyl2013prl}. This conceptual shift establishes that criticality can manifest not only as a function of control parameters, but also as a function of time in out-of-equilibrium dynamics \cite{Zvyagin2016ltp, Heyl2018rpp}. Over the past decade, the study of DQPTs has progressed from initial investigations in integrable models to encompass a broad spectrum of complex systems, including interacting \cite{Karrasch2013prb, Andraschko2014prb, Peotta2021prx}, long-range \cite{Halimeh2017prb, Stauber2017pre, Zunkovic2018prl}, non-Hermitian \cite{Zhou2018pra, Mondal2022prb}, and inhomogeneous Hamiltonians \cite{Yin2018pra, Cao2020prb, Modak2021prb, Hoyos2022prb, Kawabata2023prb}. Concurrently, the focus has shifted from establishing the mere existence of DQPTs toward elucidating universal scaling laws \cite{Heyl2015prl, Heyl2017prb, Cao2024pra} and their connections to equilibrium criticality \cite{Vajna2014prb, Heyl2018prl, Zhou2021prb, Ye2025pra}.

Parallel efforts have sought to link DQPTs with the dynamics of quantum information measures, including entanglement \cite{Jurcevic2017prl, Schmitt2018sp, Nicola2021prl, Wong2024prb}, quantum Fisher information \cite{Guan2021prr, Cheraghi2020prb, Mumford2026pra}, and OTOCs \cite{Nie2020prl, Zamani2022prb}. The phenomenology reported in these works suggests, however, that the precise relationship between DQPTs and information dynamics is system specific. For instance, in the transverse-field Ising model, the entanglement entropy of a two-site subsystem attains a maximum at DQPTs, whereas in the Su–Schrieffer–Heeger (SSH) model it exhibits only local extrema near anomalous DQPTs \cite{Wong2024prb}. A general principle governing this connection has thus far remained elusive.

In this paper, we aim to establish a universal connection between entanglement entropy and DQPTs by exploiting the momentum-space structure of translationally invariant systems. Although the Loschmidt echo is a global quantity, in systems with translational symmetry it factorizes into a product over independent momentum modes \cite{Schmitt2015prb, Vajna2015prb, Cao2025pra}, with DQPT nonanalyticities originating entirely from the vanishing of one or more momentum-resolved factors. This observation reveals that the essential physics of DQPTs is not distributed uniformly across all degrees of freedom, but is instead sharply localized within a small set of critical momentum modes \cite{Sharma2016prb, Budich2016prb}. Motivated by this, we introduce a momentum-space definition of entanglement entropy that allows us to directly probe the entanglement structure within the critical momentum subspace. We provide a comprehensive analysis of the associated information dynamics across a range of topological and superconducting systems, and we pay particular attention to the sensitivity of our results to the choice of bipartition basis. Our findings demonstrate that when the bipartition aligns with the eigenbasis of the post-quench Hamiltonian, the momentum-space entanglement entropy exhibits a robust, time-independent signature of DQPTs, thereby offering a unified geometric perspective on the interplay between entanglement, topology, and dynamical criticality.      

\section{model}
\label{sec: model}

We consider translationally invariant two-band insulators and Bogoliubov–de Gennes superconductors in one and two dimensions. Such systems are parameterized by a momentum-dependent vector $\textbf{d}_{\textbf{k}}$ via the Hamiltonian
\begin{equation}\label{eq: Hamil}
    H = \sum_{k}\textbf{c}_{\textbf{k}}^{\dagger}H_{\textbf{k}}\textbf{c}_{\textbf{k}}, H_{\textbf{k}} = \textbf{d}_{\textbf{k}} \cdot \boldsymbol{\sigma},
\end{equation}
where $\textbf{c}_{\textbf{k}}^{\dagger} = (c_{\textbf{k}, A}^{\dagger}, c_{\textbf{k}, B}^{\dagger})$ for insulators and $\textbf{c}_{\textbf{k}}^{\dagger} = (c_{\textbf{k}}^{\dagger}, c_{-\textbf{k}})$ for superconductors. The summation runs over all wave vectors in the Brillouin zone for normal insulators, and over half of the Brillouin zone for superconductors.
A sudden quench corresponds to an abrupt change in the vector field characterizing the Hamiltonian. The system is initially prepared in the ground state of $H_{\textbf{k}}^{i} = \textbf{d}_{\textbf{k}}^{i}\cdot\boldsymbol{\sigma}$, and the Hamiltonian is switched at $t=0$ to $H_{\textbf{k}}^{f} = \textbf{d}_{\textbf{k}}^{f}\cdot\boldsymbol{\sigma}$: 
\begin{equation}
    \textbf{d}_{\textbf{k}}(t) = \left\{
                                    \begin{smallmatrix} 
                                        \textbf{d}_{\textbf{k}}^{i}, & t < 0, \\ 
                                        \textbf{d}_{\textbf{k}}^{f}, & t \geq 0. 
                                    \end{smallmatrix}
    \right.
\end{equation}
Following this quench, the Loschmidt amplitude takes the compact form (see, e.g., Ref.~\cite{Vajna2015prb})
\begin{equation}
    \mathcal{G}(t) = \prod_{k}[\cos{(\varepsilon_{\textbf{k}}^{f}t)} + i\hat{\textbf{d}}_{\textbf{k}}^{i} \cdot \hat{\textbf{d}}_{\textbf{k}}^{f} \sin{(\varepsilon_{\textbf{k}}^{f}t)}],
\end{equation}
where $\hat{\textbf{d}}_{\textbf{k}}^{i(f)}$ denotes the unit vector in the direction of $\textbf{d}_{\textbf{k}}^{i(f)}$. The Fisher zeros, i.e., the solutions of $\mathcal{G}(z) = 0$ with $z = it + \tau$, are given by 
\begin{equation}\label{eq: fisher.zero}
    z_{n}(\textbf{k}) = \frac{i\pi}{\varepsilon_{\textbf{k}}^{f}}\left(n + \frac{1}{2}\right) - \frac{1}{\varepsilon_{\textbf{k}}^{f}}\arctan{[\hat{\textbf{d}}_{\textbf{k}}^{i} \cdot \hat{\textbf{d}}_{\textbf{k}}^{f}]}.
\end{equation}
A necessary condition for the occurrence of DQPTs is that the Fisher zeros approach the imaginary axis. This occurs when $\hat{\textbf{d}}_{\textbf{k}}^{i} \cdot \hat{\textbf{d}}_{\textbf{k}}^{f} = 0$ for some momentum $\textbf{k}$, i.e., when the initial and final $\textbf{d}$-vectors are perpendicular. This geometric condition establishes a connection between DQPTs and the topology of the initial and final systems, as discussed in Ref.~\cite{Vajna2015prb}. 

To elucidate the connection between entanglement entropy and DQPTs, we adopt a momentum-space definition of the entanglement entropy and examine its behavior at the critical momenta where DQPTs occur. Analogous to a real-space bipartition into subsystems $A$ and $B$, each momentum mode $\textbf{k}$ must possess at least two internal degrees of freedom. In a two-band system, these are naturally provided by the two eigenstate bands. Within a given momentum subspace, these two degrees of freedom constitute a natural bipartition, allowing the definition of basis states $|A_{\textbf{k}}\rangle$ and $|B_{\textbf{k}}\rangle$. The time-evolved state is then expressed as $|\psi_{\textbf{k}}(t)\rangle = a_{\textbf{k}}(t)|A_{\textbf{k}}\rangle + b_{\textbf{k}}(t)|B_{\textbf{k}}\rangle$, from which the full density matrix in the $\textbf{k}$ subspace follows immediately:
\begin{equation}
    \rho_{\textbf{k}}(t) = |\psi_{\textbf{k}}(t)\rangle \langle\psi_{\textbf{k}}(t)| 
    = \begin{pmatrix} 
        |a_{\textbf{k}}(t)|^{2} & a_{\textbf{k}}(t)b_{\textbf{k}}^{*}(t) \\ 
        a_{\textbf{k}}^{*}(t)b_{\textbf{k}}(t) & |b_{\textbf{k}}(t)|^{2} 
      \end{pmatrix}.
\end{equation}
Tracing out one of the two degrees of freedom yields the reduced density matrix
\begin{equation}
    \rho_{A,\textbf{k}}(t) = \operatorname{Tr}_{B}\rho_{\textbf{k}}(t), 
\end{equation}
which can be diagonalized as $\rho_{A,\textbf{k}}(t) = \operatorname{diag}{(p_{\textbf{k}}, 1-p_{\textbf{k}})}$. The associated momentum-space entanglement entropy is therefore defined as
\begin{equation}
    \begin{split}
        \mathcal{S}_{\textbf{k}}(t) & = -\text{Tr}[\rho_{A,\textbf{k}}(t)\ln{\rho_{A,\textbf{k}}(t)}] \\
        & = -p_{\textbf{k}}\ln{p_{\textbf{k}}} - (1 - p_{\textbf{k}})\ln{(1 - p_{\textbf{k}})},
    \end{split}    
\end{equation}
where the set $\{p_{\textbf{k}}, 1-p_{\textbf{k}}\}$ is referred to as the momentum-space entanglement spectrum.

\section{Benchmarks simulations}
\label{sec: simul}

We illustrate the behavior of momentum-space entanglement entropy and entanglement spectrum using three benchmark examples. 

First, we consider the SSH model. The SSH model describes a one-dimensional tight-binding chain originally introduced to model polyacetylene \cite{Su1979prl, Asboth2016}. It serves as perhaps the simplest example of a topological insulator and belongs to the BDI symmetry class \cite{Ryu2010njp}. The Hamiltonian is parameterized by the vector $\textbf{d}_{k} = (t_{1}+t_{2}\cos{k}, t_{2}\sin{k}, 0)$, where $t_{1}$ and $t_{2}$ are the staggered hopping amplitudes. The ground state is topologically trivial ($\nu = 0$) for $t_{1} > t_{2}$ and nontrivial for $t_{1} < t_{2}$. For the analysis of quench dynamics, it is convenient to introduce a polar angle $\theta_{k}$ in the $xy$-plane defined via
\begin{equation}
    \tan{\theta_{k}} = \frac{t_{2}\sin{k}}{t_{1}+t_{2}\cos{k}},
\end{equation}
so that $\hat{\textbf{d}}_{k} = \textbf{d}_{k}/|\textbf{d}_{k}| = (\cos{\theta_{k}}, \sin{\theta_{k}}, 0)$. The energy spectrum consists of two bands $\pm\varepsilon_{k} = \pm|\textbf{d}_{k}|$. For each $k$, the lower and upper band eigenstates can be expressed as linear combinations of the sublattice occupation basis states $|n_{A,k}n_{B,k}\rangle$ with $n_{A,k},n_{B,k}\in\{0,1\}$ and $n_{A,k}+n_{B,k}=1$, i.e., $|\psi_{-k}\rangle = -\sin{\frac{\theta_{k}}{2}}|10\rangle + \cos{\frac{\theta_{k}}{2}}|01\rangle$ and $|\psi_{+k}\rangle = \cos{\frac{\theta_{k}}{2}}|10\rangle + \sin{\frac{\theta_{k}}{2}}|01\rangle$.

\begin{figure}
    \centering
    \includegraphics[width=1\linewidth]{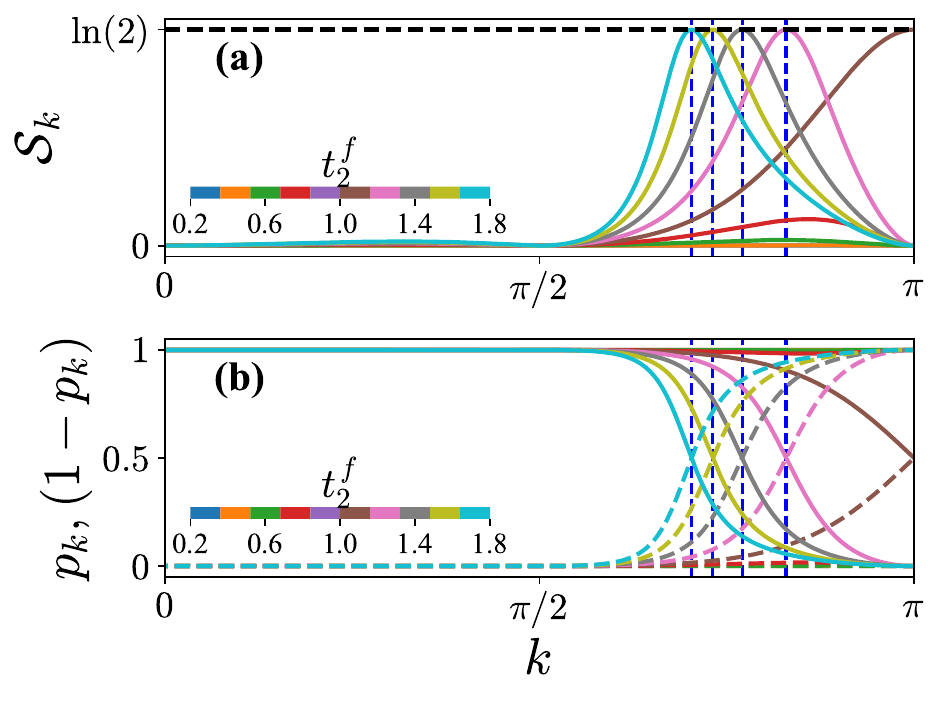}
    \caption{(a) Momentum-space entanglement entropy $\mathcal{S}_{k}$ for half of the Brillouin zone $k\in[0, \pi)$ in the SSH model, following a quench from $t_{2}^{i}=0.5$ with fixed $t_{1}=1$. (b) Corresponding entanglement spectra $p_{k}$ (solid lines) and $1-p_{k}$ (dashed lines). The vertical blue dashed lines indicate the critical momenta of DQPTs. }
    \label{fig: entanglement.ssh}
\end{figure}

In a quench protocol, the initial state is taken as the half-filled ground state corresponding to a fully occupied lower band. The time-evolved state for $t>0$ is then 
\begin{equation}\label{eq: time-state.ssh}
    |\psi_{k}(t)\rangle = \sin{\frac{\Delta\theta_{k}}{2}}e^{-i\varepsilon_{k}^{f}t}|\psi_{+,k}^{f}\rangle + \cos{\frac{\Delta\theta_{k}}{2}}e^{i\varepsilon_{k}^{f}t}|\psi_{-,k}^{f}\rangle,
\end{equation}
where $\langle\psi_{+,k}^{f}|\psi_{-,k}^{i}\rangle = \sin{\frac{\Delta\theta_{k}}{2}}$, $\langle\psi_{-,k}^{f}|\psi_{-,k}^{i}\rangle = \cos{\frac{\Delta\theta_{k}}{2}}$, and $\Delta\theta_{k} = \theta_{k}^{f} - \theta_{k}^{i}$. One readily verifies that 
\begin{equation}
    \cos{\Delta\theta_{k}} = \hat{\textbf{d}}_{k}^{i} \cdot \hat{\textbf{d}}_{k}^{f}.
\end{equation}
In the eigenbasis of the post-quench Hamiltonian $H_{k}^{f}$, each momentum mode hosts two bands (upper and lower), each of which can be either occupied or unoccupied. This structure naturally gives rise to a four-dimensional tensor-product space spanned by $\{|0^{f}0^{f}\rangle, |0^{f}1^{f}\rangle, |1^{f}0^{f}\rangle, |1^{f}1^{f}\rangle \}$. Particle number conservation (half-filling) restricts the physical subspace to states with total occupation one, so that the relevant basis reduces to $\{|0^{f}1^{f}\rangle, |1^{f}0^{f}\rangle\}$. The full density matrix in this subspace therefore reads
\begin{equation}
    \rho_{k,\text{full}}(t) = \left(\begin{smallmatrix} 
                            \cos^{2}{\frac{\Delta\theta_{k}}{2}} & \sin{\frac{\Delta\theta_{k}}{2}}\cos{\frac{\Delta\theta_{k}}{2}} \\ 
                            \sin{\frac{\Delta\theta_{k}}{2}}\cos{\frac{\Delta\theta_{k}}{2}} & \sin^{2}{\frac{\Delta\theta_{k}}{2}} 
                        \end{smallmatrix}\right).
\end{equation}   
Notably, this density matrix is independent of time $t$. Tracing out the lower band yields the reduced density matrix for the upper band:
\begin{equation}
    \rho_{k, \text{upper}} = \operatorname{Tr}_{\text{lower}}\rho_{k,\text{full}} 
               = \left(\begin{smallmatrix} 
                        \cos^{2}{\frac{\Delta\theta_{k}}{2}} & 0 \\ 
                            0 & \sin^{2}{\frac{\Delta\theta_{k}}{2}} 
                        \end{smallmatrix}
                  \right).
\end{equation}
Consequently, the momentum-space entanglement spectra are given by 
\begin{eqnarray}
    p_{k} &=& \cos^{2}{\frac{\Delta\theta_{k}}{2}} = \frac{1+\hat{\textbf{d}}_{k}^{i} \cdot \hat{\textbf{d}}_{k}^{f}}{2}, \\
    1-p_{k} &=& \sin^{2}{\frac{\Delta\theta_{k}}{2}} = \frac{1-\hat{\textbf{d}}_{k}^{i} \cdot \hat{\textbf{d}}_{k}^{f}}{2}.
\end{eqnarray}
The momentum-space entanglement entropy takes the form
\begin{equation}\label{eq: ee.ssh}
    \begin{split}
        \mathcal{S}_{k}(t) & = -\frac{1+\hat{\textbf{d}}_{k}^{i} \cdot \hat{\textbf{d}}_{k}^{f}}{2}\ln{\frac{1+\hat{\textbf{d}}_{k}^{i} \cdot \hat{\textbf{d}}_{k}^{f}}{2}} \\
        & \quad\quad - \frac{1-\hat{\textbf{d}}_{k}^{i} \cdot \hat{\textbf{d}}_{k}^{f}}{2}\ln{\frac{1-\hat{\textbf{d}}_{k}^{i} \cdot \hat{\textbf{d}}_{k}^{f}}{2}}.
    \end{split}
\end{equation}

Fig.~\ref{fig: entanglement.ssh} displays the momentum-space entanglement entropy $\mathcal{S}_{k}$ and entanglement spectrum for several quench trajectories in the SSH model. Owing to the symmetry $\mathcal{S}_{k} = \mathcal{S}_{-k}$ and $p_{k} = p_{-k}$, we restrict the plot to half of the Brillouin zone ($k > 0$). In the quench protocol, we fix $t_{1} = 1$; a DQPT occurs when the initial and final values of $t_{2}$ lie on opposite sides of the topological critical point $t_{2c}=1$. As seen in Fig.~\ref{fig: entanglement.ssh}~(a), $\mathcal{S}_{k}$ attains its maximum at the critical momenta $\pm k^{*}$. For a two-band system ($I=2$), the maximal entanglement entropy saturates the bound $\mathcal{S}_{k,\text{max}} = \ln I$ \cite{Vidal2003prl}. Simultaneously, the entanglement spectrum becomes degenerate at the critical momentum [see Fig.~\ref{fig: entanglement.ssh}~(b)], i.e., a level crossing occurs such that $p_{k^{*}} = 1 - p_{k^{*}} = 1/2$. In the special case $t_{2}^{f} = t_{2c} = 1$, the critical momentum is located at the Brillouin zone boundary $k^{*} = \pm \pi$.

\begin{figure}
    \centering
    \includegraphics[width=1\linewidth]{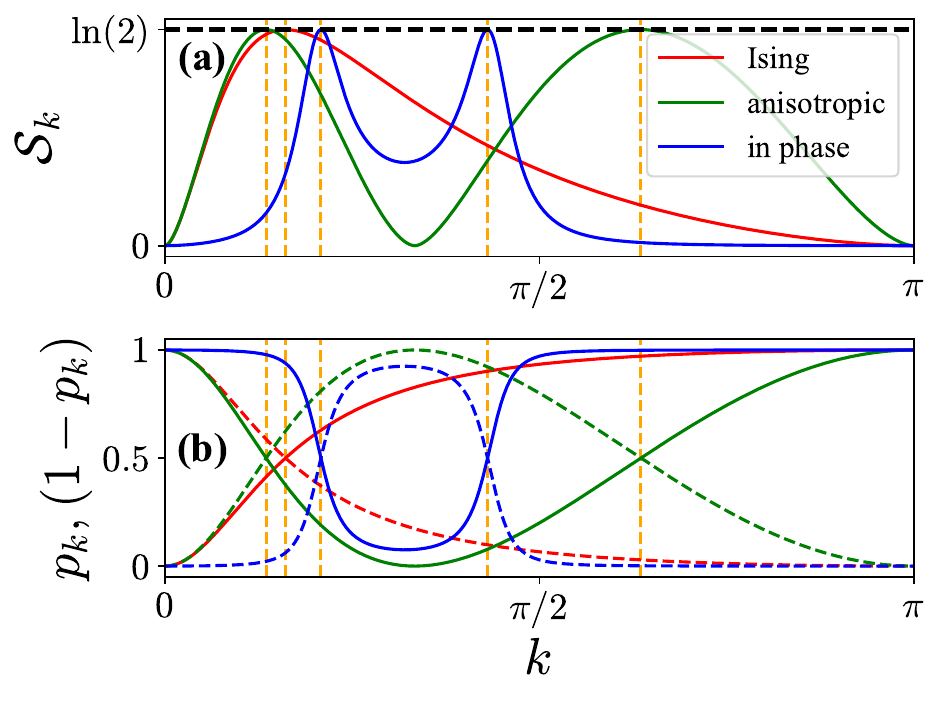}
    \caption{(a) Momentum-space entanglement entropy $\mathcal{S}_{k}$ in the quantum XY chain. (b) Associated entanglement spectra $p_{k}$ (solid lines) and $1-p_{k}$ (dashed lines). The red curve corresponds to a quench across the Ising transition from $(h_{0}, \gamma_{0}) = (0.5, 1)$ to $(h_{1}, \gamma_{1}) = (1.5, 1)$; the green curve crosses the anisotropic transition from $(h_{0}, \gamma_{0}) = (0.5, 1)$ to $(h_{1}, \gamma_{1}) = (0.5, -1)$; and the blue curve lies entirely within the FM$_{x}$ phase, from $(h_{0}, \gamma_{0}) = (0.2, 0.1)$ to $(h_{1}, \gamma_{1}) = (0.8, 0.1)$. The orange dashed vertical lines indicate the critical momenta of DQPTs.}
    \label{fig: entanglement.xy}
\end{figure}

\begin{figure*}
    \centering
    \includegraphics[width=1\linewidth]{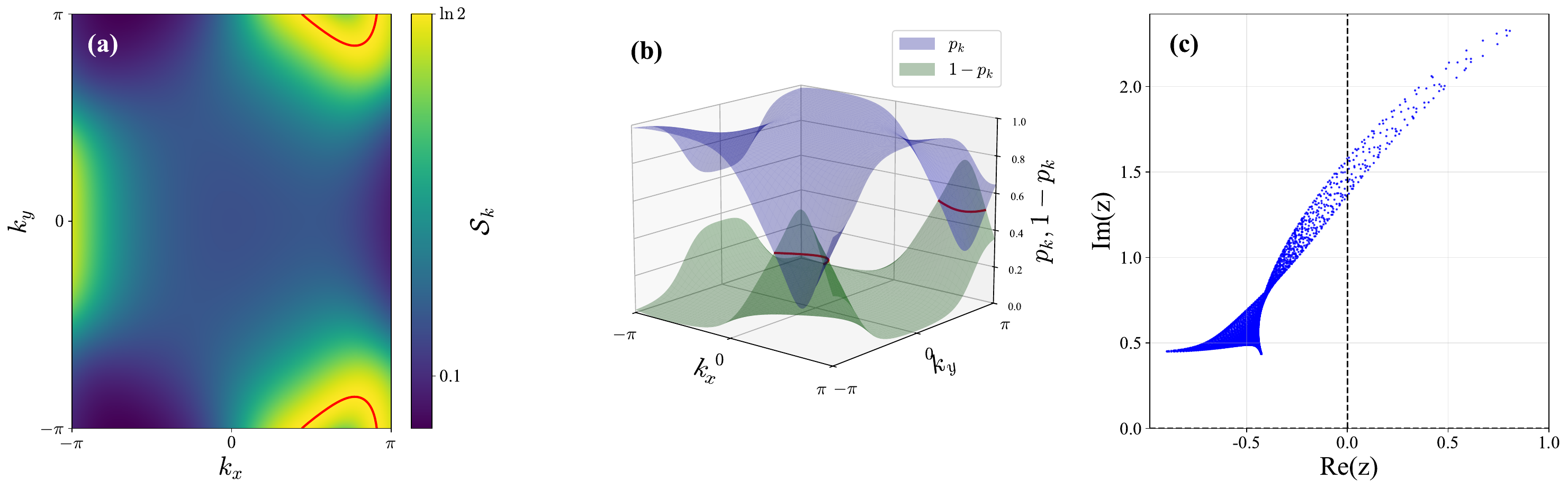}
    \caption{(a) Contour plot of the momentum-space entanglement entropy $\mathcal{S}_{\textbf{k}}$ for the Haldane model with $\phi = \frac{\pi}{2}$, $\gamma_{1} = 1$, and $\gamma_{2} = 0.3$. The quench is performed from $m_{i} = 0.5$ to $m_{f} = 2$; the critical point separating the trivial and topological phases is $m_{c} = 3\sqrt{3}\gamma_{2} \approx 1.56$. The red curves indicate the momenta $\textbf{k}$ at which the entanglement entropy attains its maximal value $\mathcal{S}_{\textbf{k}^{*}} = \ln{2}$. (b) Three-dimensional representation of the entanglement spectra $p_{\textbf{k}}$ and $1-p_{\textbf{k}}$. The red lines denote the critical momenta satisfying $p_{\textbf{k}} = 1 - p_{\textbf{k}}$. (c) Fisher zeros $z_{0}$ in the complex time plane.}
    \label{fig: e.haldane}
\end{figure*}

As a second benchmark, we consider one-dimensional superconductors, exemplified by the Kitaev chain \cite{kitaev2001pu}. In fact, one-dimensional quantum spin systems also fall into this category, including the transverse-field Ising model and the quantum XY chain. These spin models can be mapped to spinless fermions via the Jordan–Wigner transformation and share the same Bogoliubov–de Gennes Hamiltonian form as the Kitaev chain \cite{Suzuki2013}. Owing to its relatively rich and well-established quantum phase diagram, we focus on the quantum XY chain as our second example. After applying the Jordan–Wigner transformation, the quantum XY chain becomes equivalent to a one-dimensional $p$-wave Kitaev model and is parameterized by the vector $\textbf{d}_{k} = (0, -\gamma\sin{k}, h - \cos{k})$. Here, $\gamma$ denotes the anisotropy parameter and $h$ the external magnetic field. The system undergoes an Ising transition at $h_{c} = 1$ and an anisotropic transition at $\gamma_{c} = 0$. Diagonalization of the Hamiltonian (\ref{eq: Hamil}) is accomplished via a Bogoliubov transformation, $c_{k} = u_{k}\eta_{k} + iv_{k}\eta_{-k}^{\dagger}$. Defining the Bogoliubov angle $\theta_{k}$ via $\tan{2\theta_{k}} = \gamma\sin{k}/(h - \cos{k})$, the coefficients are expressed as $u_{k} = \cos{\theta_{k}}$ and $v_{k} = \sin{\theta_{k}}$.

Consider a quench from $H_{k}^{i} = H(h_{0}, \gamma_{0})$ to $H_{k}^{f} = H(h_{1}, \gamma_{1})$. For each $k > 0$, the initial state can be written as $|\psi_{k}(0)\rangle = \cos{\Delta\theta_{k}}|0_{k}^{f}0_{-k}^{f}\rangle + \sin{\Delta\theta_{k}}|1_{k}^{f}1_{-k}^{f}\rangle$, where $|0_{k}^{f}0_{-k}^{f}\rangle$ and $|1_{k}^{f}1_{-k}^{f}\rangle = \eta_{k}^{f\dagger}\eta_{-k}^{f\dagger}|0_{k}^{f}0_{-k}^{f}\rangle$ are eigenstates of the post-quench Hamiltonian $\tilde{H}_{k}$, and $\Delta\theta_{k} = \theta_{k}^{f} - \theta_{k}^{i}$. The time-evolved state is therefore given by \cite{Cao2024pra}
\begin{equation}
    |\psi_{k}(t)\rangle = \cos{\Delta\theta_{k}}e^{i\varepsilon_{k}^{f}t}|0_{k}^{f}0_{-k}^{f}\rangle + \sin{\Delta\theta_{k}}e^{-i\varepsilon_{k}^{f}t}|1_{k}^{f}1_{-k}^{f}\rangle,
\end{equation}
where $\varepsilon_{k}^{f} = |\textbf{d}_{k}^{f}| = \sqrt{(h_{1}-\cos{k})^{2} + \gamma_{1}^{2}\sin^{2}{k}}$ and $\cos{\Delta\theta_{k}} = \hat{\textbf{d}}_{k}^{i}\cdot\hat{\textbf{d}}_{k}^{f}$. In contrast to the SSH model, the full density matrix $\rho_{k,-k}$ here is defined in the two-dimensional basis $\{|0_{k}^{f}0_{-k}^{f}\rangle, |1_{k}^{f}1_{-k}^{f}\rangle\}$, yielding
\begin{equation}
    \rho_{k, -k}(t) = \left(\begin{smallmatrix} 
                            \cos^{2}{\Delta\theta_{k}} & \sin{\Delta\theta_{k}}\cos{\Delta\theta_{k}} \\ 
                            \sin{\Delta\theta_{k}}\cos{\Delta\theta_{k}} & \sin^{2}{\Delta\theta_{k}} 
                        \end{smallmatrix}\right),
\end{equation}
which, as before, is independent of time. The reduced density matrix for the $k$-mode is obtained by tracing out its partner $-k$:
\begin{equation}
    \rho_{k} = \operatorname{Tr}_{-k}\rho_{k,-k} = \operatorname{diag}(\cos^{2}{\Delta\theta_{k}}, \sin^{2}{\Delta\theta_{k}}),
\end{equation}
where
\begin{equation}
    p_{k} = \cos^{2}{\Delta\theta_{k}} = \frac{1 + \hat{\textbf{d}}_{k}^{i}\cdot\hat{\textbf{d}}_{k}^{f}}{2}.
\end{equation}
We thus verify that the expression for the momentum-space entanglement entropy in the superconducting case coincides with that of the SSH model, as given by Eq.~(\ref{eq: ee.ssh}).

Fig.~\ref{fig: entanglement.xy} displays the momentum-space entanglement entropy $\mathcal{S}_{k}$ and entanglement spectrum for several quench trajectories in the quantum XY chain. The behavior mirrors that observed in the SSH model: $\mathcal{S}_{k}$ attains its maximum value $\ln 2$ at the critical momenta, and the entanglement spectrum exhibits a level crossing at the same $k^{*}$. This holds true even when multiple critical momenta appear, as in quenches crossing the anisotropic transition or those exhibiting accidental DQPTs within a single phase \cite{Vajna2014prb}.

As a third benchmark, we consider a two-dimensional system: the Haldane model on a honeycomb lattice \cite{Haldane1988prl}. The Bloch Hamiltonian of this model is given by $\mathcal{H}_{\textbf{k}} = \textbf{d}_{\textbf{k}} \cdot \boldsymbol{\sigma} = d_{x\textbf{k}}\sigma_{x} + d_{y\textbf{k}}\sigma_{y} + d_{z\textbf{k}}\sigma_{z}$, with
\begin{equation}
    d_{x\textbf{k}} = \gamma_{1}\sum_{j=1}^{3}\cos{(\textbf{k} \cdot \textbf{a}_{j})}, d_{y\textbf{k}} = \gamma_{1}\sum_{j=1}^{3}\sin{(\textbf{k} \cdot \textbf{a}_{j})},
\end{equation}
\begin{equation}
    d_{z\textbf{k}} = m - 2\gamma_{2}\sin{\phi}\sum_{j=1}^{3}\cos{(\textbf{k} \cdot \textbf{b}_{j})},
\end{equation}
where the vectors $\textbf{a}_{j}$ and $\textbf{b}_{j}$ connect nearest-neighbor and next-nearest-neighbor lattice sites, respectively. The Chern number depends on the phase $\phi$, the next-nearest-neighbor hopping amplitude $\gamma_{2}$, and the mass term $m$: for $|m| > |3\sqrt{3}\gamma_{2}\sin{\phi}|$, the system is a trivial insulator with Chern number $Q = 0$, whereas for $|m| < |3\sqrt{3}\gamma_{2}\sin{\phi}|$, it is a topological insulator with $Q = \pm 1$. The eigenstates of the Hamiltonian are given by
\begin{equation}
    |\psi_{+,\textbf{k}}\rangle = \left(\begin{smallmatrix} 
                                            \cos{\frac{\theta_{\textbf{k}}}{2}} \\ 
                                            \sin{\frac{\theta_{\textbf{k}}}{2}}e^{i\varphi_{\textbf{k}}}  
                                        \end{smallmatrix}\right),
    |\psi_{-,\textbf{k}}\rangle = \left(\begin{smallmatrix} 
                                            -\sin{\frac{\theta_{\textbf{k}}}{2}} \\ 
                                            \cos{\frac{\theta_{\textbf{k}}}{2}}e^{i\varphi_{\textbf{k}}}  
                                        \end{smallmatrix}\right),                        
\end{equation}
where the angles $\theta_{\textbf{k}}$ and $\varphi_{\textbf{k}}$ are determined by the spherical parametrization of $\textbf{d}_{\textbf{k}}$: $d_{x,\textbf{k}} = |\textbf{d}_{\textbf{k}}|\sin{\theta_{\textbf{k}}}\cos{\varphi_{\textbf{k}}}$, $d_{y,\textbf{k}} = |\textbf{d}_{\textbf{k}}|\sin{\theta_{\textbf{k}}}\sin{\varphi_{\textbf{k}}}$, and $d_{z,\textbf{k}} = |\textbf{d}_{\textbf{k}}|\cos{\theta_{\textbf{k}}}$.

For a quench from $\textbf{d}_{\textbf{k}}^{i}$ to $\textbf{d}_{\textbf{k}}^{f}$, the time-evolved state takes the form
\begin{equation}
    |\psi_{\textbf{k}}(t)\rangle = \cos{\frac{\Delta\phi_{\textbf{k}}}{2}}e^{-i\varepsilon_{\textbf{k}}^{f}t} |\psi_{+\textbf{k}}^{f}\rangle + \sin{\frac{\Delta \phi_{\textbf{k}}}{2}}e^{i\varepsilon_{\textbf{k}}^{f}t} |\psi_{-\textbf{k}}^{f}\rangle,
\end{equation}
where $\cos{\Delta\phi_{\textbf{k}}} = \hat{\textbf{d}}_{\textbf{k}}^{i}\cdot\hat{\textbf{d}}_{\textbf{k}}^{f}$. Analogous to the SSH model, the full density matrix in the eigenbasis of the post-quench Hamiltonian $H_{\textbf{k}}^{f}$ reads
\begin{equation}
    \rho_{\textbf{k}, \text{full}} = \left(\begin{smallmatrix} 
                                                \cos^{2}{\frac{\Delta\phi_{\textbf{k}}}{2}} & \sin{\frac{\Delta\phi_{\textbf{k}}}{2}}\cos{\frac{\Delta\phi_{\textbf{k}}}{2}} \\ 
                                                \sin{\frac{\Delta\phi_{\textbf{k}}}{2}}\cos{\frac{\Delta\phi_{\textbf{k}}}{2}} & \sin^{2}{\frac{\Delta\phi_{\textbf{k}}}{2}} 
                                            \end{smallmatrix}\right),
\end{equation}
The reduced density matrix for the upper band is then obtained by tracing out the lower band:
\begin{equation}
    \rho_{\textbf{k}, \text{upper}} = \operatorname{Tr}_{\text{lower}}\rho_{\textbf{k},\text{full}} = \left(\begin{smallmatrix} 
                                                \cos^{2}{\frac{\Delta\phi_{\textbf{k}}}{2}} & 0 \\ 
                                                0 & \sin^{2}{\frac{\Delta\phi_{\textbf{k}}}{2}} 
                                            \end{smallmatrix}\right).
\end{equation}
Consequently, the momentum-space entanglement entropy in the Haldane model is also given by Eq.~(\ref{eq: ee.ssh}).

Fig.~\ref{fig: e.haldane}~(a) and (b) display the momentum-space entanglement entropy $\mathcal{S}_{\textbf{k}}$ and the corresponding entanglement spectrum for the Haldane model with representative parameters $\phi = \frac{\pi}{2}$, $\gamma_{1} = 1$, and $\gamma_{2} = 0.3$, for which the critical point between the trivial and topological phases is $m_{c} = 3\sqrt{3}\gamma_{2} \approx 1.56$. In contrast to the one-dimensional case, the critical momenta $\textbf{k}^{*}$ here form continuous lines along which the entanglement entropy $\mathcal{S}_{\textbf{k}}$ attains its maximal value $\ln 2$ and the entanglement spectrum becomes degenerate. This distinction reflects a fundamental difference between DQPTs in one and two dimensions: in one dimension the critical momenta typically constitute isolated points (or lines in parameter space) \cite{Heyl2013prl, Heyl2018rpp}, whereas in two dimensions they span extended regions [see Fig.~\ref{fig: e.haldane}~(c)].

\section{conclusion}

In this work, we have systematically investigated the critical behavior of momentum-space entanglement entropy at DQPTs across a variety of translationally invariant systems. Through detailed analysis of three representative benchmarks---the one-dimensional SSH model, the one-dimensional quantum XY chain (mapped to a Kitaev chain), and the two-dimensional Haldane model---we have established a universal connection between the geometric conditions governing DQPTs and the spectral properties of the reduced density matrix in momentum space.

Our analysis reveals that the necessary condition for the occurrence of DQPTs, namely the orthogonality of the initial and final Hamiltonian vectors $\hat{\textbf{d}}_{\textbf{k}}^{i} \cdot \hat{\textbf{d}}_{\textbf{k}}^{f} = 0$, directly dictates the behavior of the entanglement spectrum when the bipartition is chosen to align with the eigenbasis of the post-quench Hamiltonian. In all three models examined, we find that at the critical momenta $\textbf{k}^{*}$ where the Fisher zeros approach the imaginary axis, the momentum-space entanglement spectrum $\{p_{\textbf{k}}, 1-p_{\textbf{k}}\}$ becomes exactly degenerate, i.e., $p_{\textbf{k}^{*}} = 1-p_{\textbf{k}^{*}} = 1/2$. This degeneracy corresponds precisely to the configuration of maximal entanglement entropy, with $\mathcal{S}_{\textbf{k}^{*}}$ saturating the theoretical upper bound $\ln 2$ for a two-band system.

The nature of this critical structure exhibits a distinct dependence on spatial dimensionality. In one-dimensional systems, such as the SSH model and the quantum XY chain, the critical momenta appear as isolated pairs of points in the Brillouin zone. In contrast, for the two-dimensional Haldane model, the condition $\hat{\textbf{d}}_{\textbf{k}}^{i} \cdot \hat{\textbf{d}}_{\textbf{k}}^{f} = 0$ defines continuous curves in momentum space. Consequently, the degenerate entanglement spectrum and maximal entropy are observed along entire extended one-dimensional manifolds within the two-dimensional Brillouin zone. This distinction underscores the dimensional dependence of DQPTs: while they manifest as discrete temporal cusps associated with isolated $\textbf{k}$-modes in 1D, they form continuous families of critical modes in higher dimensions.

Crucially, we have demonstrated that the connection between entanglement entropy and DQPTs is highly sensitive to the choice of the bipartition basis. As shown in the Appendix for the SSH model, selecting the sublattice basis $(A, B)$---rather than the eigenbasis of the post-quench Hamiltonian---yields a qualitatively different behavior. In the sublattice basis, the momentum-space entanglement entropy at the critical momentum $\mathcal{S}_{k^{*}}(t)$ becomes explicitly time-dependent and attains a \emph{minimum} precisely at the critical times $t_{n}$ of the DQPTs, in stark contrast to the time-independent maximal entropy observed in the eigenbasis. Moreover, the times at which the sublattice entropy reaches its maximum are shifted to half of the DQPT critical times. This dichotomy highlights that the geometric DQPT condition $\hat{\textbf{d}}_{\textbf{k}}^{i} \cdot \hat{\textbf{d}}_{\textbf{k}}^{f} = 0$ does not universally guarantee maximal entanglement; rather, it is the alignment of the bipartition with the natural eigenmodes of the post-quench dynamics that reveals the most direct signature of the underlying dynamical criticality.

Our findings elucidate the intrinsic link between the non-analytic temporal evolution of the Loschmidt echo and the static, time-independent structure of momentum-space correlations when viewed in the appropriate basis. The fact that the reduced density matrix $\rho_{\textbf{k}}$ in the eigenbasis remains explicitly time-independent, even as the full state undergoes non-trivial unitary evolution, underscores that the entanglement signatures of DQPTs are encoded entirely in the geometrical mismatch between the initial and final topological configurations. This framework provides a robust, computationally accessible diagnostic for identifying and classifying DQPTs, while simultaneously cautioning that the choice of bipartition must be carefully considered. Future work may extend this momentum-space entanglement perspective to periodically driven (Floquet) systems, interacting settings, or to the characterization of higher-order dynamical criticalities, where the interplay between basis dependence and criticality promises to be even richer.

\begin{acknowledgments}
  K.C. was funded by Basic Research Program of Jiangsu (Grant No.~BK20250886) and Basic Research Program of Yangzhou (Grant No.~YZ2025132). J.W. was supported by the National Natural Science Foundation of China (Grant No.~11875047). S.C. was supported by National Key Research and Development Program of China (Grant No.~2021YFA1402104) and the National Natural Science Foundation under Grants No.~12474287 and No.~T2121001.
\end{acknowledgments}

\appendix

\section{Momentum-space entanglement entropy for the sublattice in insulators}

The definition of entanglement entropy depends sensitively on the choice of bipartition. In the main text, we demonstrated that the entanglement entropy defined with respect to the eigenbasis of the post-quench Hamiltonian exhibits a robust connection with DQPTs. A natural question then arises: what behavior emerges if an alternative bipartition basis is selected? Here, we explore this question using the sublattice basis $(A, B)$ of the SSH model as an illustrative example.

\begin{figure}
    \centering
    \includegraphics[width=1\linewidth]{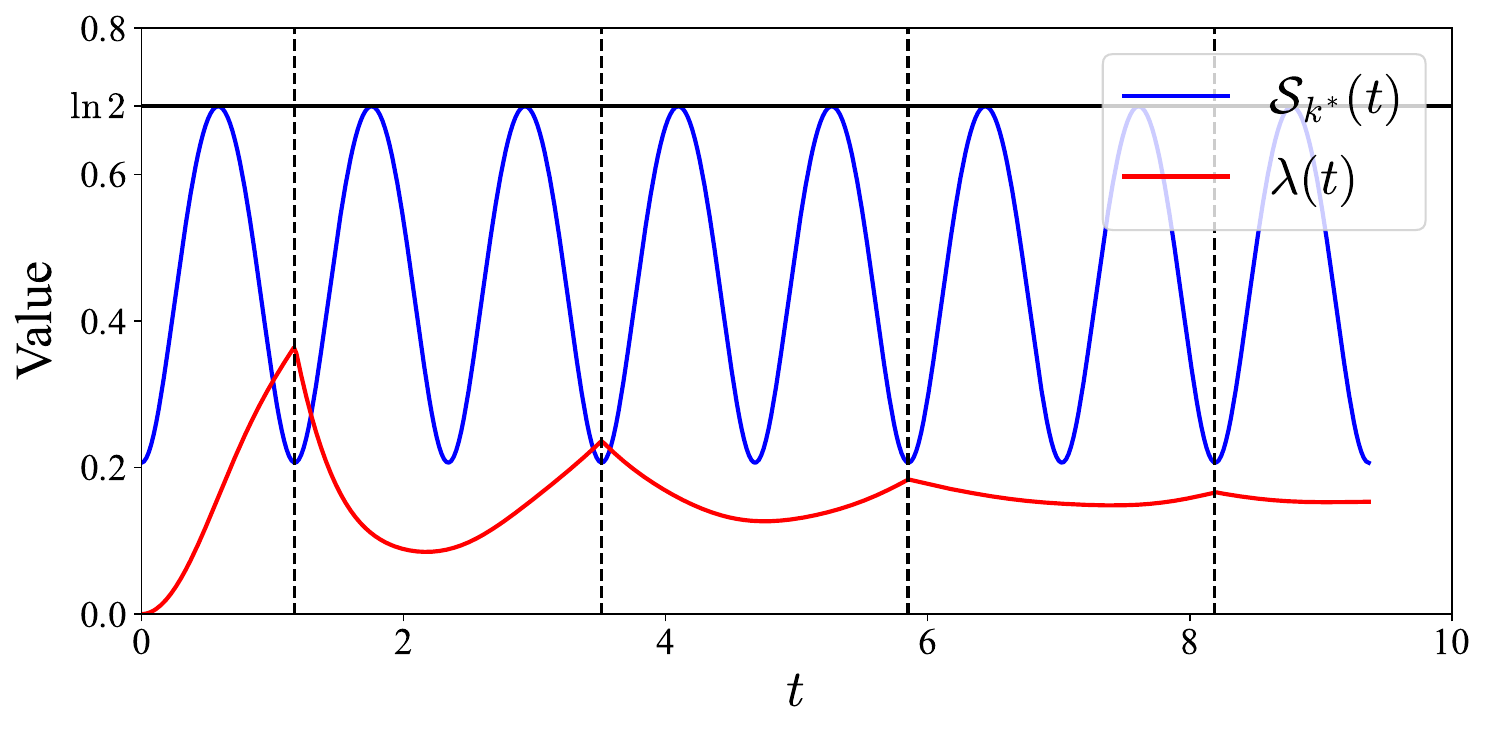}
    \caption{Momentum-space entanglement entropy $\mathcal{S}_{k^{*}}(t)$ evaluated at the critical momentum $k^{*} \approx 2.50$ in the sublattice basis $(A, B)$. The quench is performed from $t_{2}^{i} = 0.5$ to $t_{2}^{f} = 2.0$ with fixed $t_{1} = 1$. The rate function $\lambda(t)$ is shown for reference, with critical times of DQPTs indicated by vertical dashed lines.}
    \label{fig: sublattice.ssh}
\end{figure}

From Eq.~(\ref{eq: time-state.ssh}), the time-evolved state expressed in the sublattice basis follows immediately:
\begin{equation}
    \begin{split}
        |\psi_{k}(t) &= (\sin{\frac{\Delta \theta_{k}}{2}}e^{-i\varepsilon_{k}^{f}t}\cos{\frac{\theta_{k}^{f}}{2}} - \cos{\frac{\Delta \theta_{k}}{2}}e^{i\varepsilon_{k}^{f}t}\sin{\frac{\theta_{k}^{f}}{2}})|A\rangle \\
        & + (\sin{\frac{\Delta \theta_{k}}{2}}e^{-i\varepsilon_{k}^{f}t}\sin{\frac{\theta_{k}^{f}}{2}} - \cos{\frac{\Delta \theta_{k}}{2}}e^{i\varepsilon_{k}^{f}t}\cos{\frac{\theta_{k}^{f}}{2}})|B\rangle,
    \end{split}
\end{equation}
where $|A\rangle = |1_{A}0_{B}\rangle$ and $|B\rangle = |0_{A}1_{B}\rangle$. For brevity, we write $|\psi_{k}(t)\rangle = a_{k}(t)|A\rangle + b_{k}(t)|B\rangle$. The full density matrix in the sublattice basis $(A, B)$ is then
\begin{equation}
    \rho_{k}(t) = \left(\begin{smallmatrix} 
                            |a_{k}(t)|^{2} & a_{k}(t)b_{k}^{*}(t) \\ 
                            a_{k}^{*}(t)b_{k}(t) & |b_{k}(t)|^{2} 
                        \end{smallmatrix}\right).
\end{equation}
Tracing out sublattice $B$ yields the reduced density matrix for sublattice $A$:
\begin{equation}
    \rho_{A,k}(t) = \operatorname{Tr}_{B}\rho_{k}(t) = \left(\begin{smallmatrix} 
                                                                |a_{k}(t)|^{2} & 0 \\ 
                                                                0 &  |b_{k}(t)|^{2}
                                                            \end{smallmatrix}\right),
\end{equation}
where the time-dependent occupation probability is given by
\begin{equation}
    |a_{k}(t)|^{2} = \frac{1}{2} - \frac{1}{2}[\cos{\Delta\theta_{k}}\cos{\theta_{k}^{f}} + \sin{\Delta\theta_{k}}\sin{\theta_{k}^{f}}\cos{2\varepsilon_{k}^{f}t}].
\end{equation}
The corresponding momentum-space entanglement entropy is therefore
\begin{equation}
    \mathcal{S}_{k}(t) = -|a_{k}(t)|^{2}\ln{|a_{k}(t)|^{2}} - (1 - |a_{k}(t)|^{2})\ln{(1 - |a_{k}(t)|^{2})}.
\end{equation}
The entropy attains its maximal value $\ln 2$ when $|a_{k}(t)|^{2} = 1/2$. This condition requires both $\cos{\Delta\theta_{k}} = \hat{\textbf{d}}_{k}^{i}\cdot\hat{\textbf{d}}_{k}^{f} = 0$ (i.e., the DQPT geometric condition) and $\cos{2\varepsilon_{k}^{f}t} = 0$, which together imply
\begin{equation}
    t = \frac{\pi}{2\varepsilon_{k}^{f}}(n + \frac{1}{2}), n\in\mathbb{Z}.
\end{equation}
Recalling from Eq.~(\ref{eq: fisher.zero}) that the critical times of DQPTs are $t_{n} = \frac{\pi}{\varepsilon_{k}^{f}}(n + \frac{1}{2})$, we observe that the maxima of the sublattice entanglement entropy occur precisely at half of the DQPT critical times:
\begin{equation}
    t_{\text{max, entropy}} = \frac{1}{2}t_{\text{DQPT}}.
\end{equation}

Fig.~\ref{fig: sublattice.ssh} displays the rate function $\lambda(t)$ alongside the momentum-space entanglement entropy $\mathcal{S}_{k}(t)$ evaluated in the sublattice basis at the critical momentum $k^{*}$ for the representative quench. In stark contrast to the behavior observed in the eigenbasis bipartition, the entanglement entropy $\mathcal{S}_{k^{*}}(t)$ here attains a \emph{minimum}---rather than a maximum---at the critical times of the DQPTs.

\bibliography{reference}

\end{document}